\documentclass[a4paper,11pt]{article}

\usepackage{fullpage}

\usepackage[utf8]{inputenc}
\usepackage[T1]{fontenc}

\usepackage{microtype} %

\usepackage[compact]{titlesec}

\newcommand{\shrinkitems}{\setlength{\itemsep}{0ex}} %
\newcommand{\shrinktable}{\setlength{\abovecaptionskip}{-1ex}\setlength{\belowcaptionskip}{-1ex}} %

\usepackage[usenames]{xcolor}
\usepackage[bf, small]{caption}

\newcommand{\ttlpar}[1]{\noindent\textbf{#1}}

\usepackage{bold-extra}
\usepackage[final]{listings}
\usepackage{courier}
\usepackage{booktabs}
\usepackage{multirow}
\usepackage[pdfborder={0 0 0}]{hyperref}
\usepackage{balance}

\usepackage{tikz}
\usetikzlibrary{shapes}
\usetikzlibrary{matrix}
\usetikzlibrary{positioning}

\usepackage{enumerate}
\usepackage{amsmath}
\usepackage[amsmath, thmmarks]{ntheorem}

\theorembodyfont{\upshape}
\theoremheaderfont{\sc}
\newtheorem{theorem}{Theorem}[section]
\newtheorem{lemma}[theorem]{Lemma}

\newtheorem{observation}[theorem]{Observation}
\theoremheaderfont{\it}
\theoremstyle{nonumberplain}
\theoremseparator{}
\theoremsymbol{\rule{1ex}{1ex}}
\newtheorem{Proof}{Proof}

\DeclareMathOperator{\Access}{Access}
\DeclareMathOperator{\Rank}{Rank}
\DeclareMathOperator{\Excess}{Excess}
\DeclareMathOperator{\Select}{Select}

\DeclareMathOperator{\Lookup}{Lookup}
\DeclareMathOperator{\fwdsearch}{FwdSearch}
\DeclareMathOperator{\FindClose}{FindClose}
\DeclareMathOperator{\FindOpen}{FindOpen}

\newcommand{\bit}[1]{\text{\bf\texttt{#1}}}
\newcommand{\bitnobf}[1]{\text{\texttt{#1}}}
\newcommand{\bitzero}{\bit{0}}
\newcommand{\bitone}{\bit{1}}
\newcommand{\paropen}{\text{\bf\texttt{(}}}
\newcommand{\parclose}{\text{\bf\texttt{)}}}
\newcommand{\Lhigh}{L^\text{high}}
\newcommand{\Llow}{L^\text{low}}

\title{Fast Compressed Tries through Path Decompositions}
\author{
  Roberto Grossi \\ \small Università di Pisa \\ \small \texttt{grossi@di.unipi.it}
  \and
  Giuseppe Ottaviano\footnote{Part of the work done while the author was an intern at Microsoft Research, Cambridge.} \\ \small Università di Pisa \\ \small \texttt{ottavian@di.unipi.it}
}

\date{}

\begin{document}
\maketitle

\begin{abstract}
  Tries are popular data structures for storing a set of strings,
  where common prefixes are represented by common root-to-node
  paths. Over fifty years of usage have produced many variants and
  implementations to overcome some of their limitations. We explore
  new succinct representations of path-decomposed tries and
  experimentally evaluate the corresponding reduction in space usage
  and memory latency, comparing with the state of the art. We study
  two cases of applications: \emph{(1)}~a compressed dictionary for
  (compressed) strings, and \emph{(2)}~a monotone minimal perfect hash
  for strings that preserves their lexicographic order.

  For~\emph{(1)}, we obtain data structures that outperform other
  state-of-the-art compressed dictionaries in space efficiency, while
  obtaining predictable query times that are competitive with data
  structures preferred by the practitioners. In~\emph{(2)}, our tries
  perform several times faster than other trie-based monotone perfect
  hash functions, while occupying nearly the same space.
\end{abstract}

\setcounter{page}{0}
\thispagestyle{empty}

\newpage

\section{Introduction}
\label{sec:introduction}

Tries are a widely used data structure that turn a string set into a
digital search tree. Several operations can be supported, such as mapping the
strings to integers, re\emph{trie}ving a string from the trie, performing
prefix searches, and many others.
Thanks to their simplicity and functionality, they have enjoyed a
remarkable popularity in a number of fields---Computational Biology, Data
Compression, Data Mining, Information Retrieval, Natural Language
Processing, Network Routing, Pattern Matching, Text Processing, and
Web applications, to name a few---motivating the significant effort
spent in the variety of their implementations over the last fifty~years.

However their simplicity comes at a cost: as most tree structures,
they generally suffer poor locality of reference due to
pointer-chasing. This effect is amplified when using space efficient
representations of tries, where performing any basic navigational
operation, such as visiting a child, requires accessing possibly
several directories, usually with unpredictable memory access
patterns. Tries are particularly affected as they are unbalanced
structures: the height can be in the order of the number of strings in
the set. Furthermore, space savings are achieved only by
exploiting the common prefixes in the string set, while it is not clear
how to compress their nodes and their labels without incurring an
unreasonable overhead in the running time.

In this paper, we experiment how \emph{path decompositions} of tries
help on both the above mentioned issues, inspired by the work
presented in \cite{pods08}.  By using a \emph{centroid} path
decomposition, the height is guaranteed to be logarithmic, reducing
dramatically the number of cache misses in a traversal; besides, for
any path decomposition the labels can be laid out in a way that
enables efficient compression and decompression of a label in a
sequential fashion.

We keep two main goals in mind: $(i)$~reduce the space requirement,
and $(ii)$~guarantee fast query times using algorithms that exploit
the memory hierarchy. In our algorithm engineering design, we follow
some guidelines: $(a)$~the proposed algorithms and data structures
should be the simplest possible to ensure reproducibility of the
results, while the performance should be similar to or better than
what is available in the state of the art. $(b)$~The proposed
techniques should possibly lay on a theoretical ground. $(c)$~The theoretical
complexity of some operations is allowed to be worse than that known
for the best solutions when there is a clear experimental
benefit\footnote{For example, it is folklore that a sequential scan of
  a \emph{small} sorted set of keys is faster than a binary search
  because the former method is very friendly with branch prediction
  and cache pre-fetching of modern machines.}, since we seek for the
best performance in practice.

The literature about space-efficient and cache-efficient tries is
vast. Several papers address the issue of a cache-friendly access to a
set of strings supporting prefix search, e.g.\mbox{}
\cite{AcharyaZS99,BenderFK06,BrodalF06,FerraginaGrossi99} but they do
not deal with space issues except \cite{BenderFK06}, which introduces
an elegant variant of front coding. Other papers aiming at succinct
labeled trees and compressed data structures for strings, e.g.\mbox{}
\cite{bpalx10,monotonehash09,dfuds,BlandfordB08,FerraginaLMM09,Munro:2001:SRB,bpsoda10},
support powerful operations---such as path queries---and are very good
in compressing data, but they do not exploit the memory hierarchy. Few
papers \cite{ChiuHSV10,pods08} combine (nearly) optimal information
theoretic bounds for space occupancy with good cache efficient
bounds, but no experimental analysis is performed. More references on compressed string dictionaries can be found
in~\cite{csd11}.

The paper is organized as follows. We apply our path decomposition ideas to string dictionaries in
Section~\ref{sec:stringdict} and to monotone perfect hash functions
(hollow tries) in Section~\ref{sec:hollow}, showing that it is
possible to improve their performance with a very small
space overhead. In Section~\ref{sec:rmtree}, we present some
optimizations to the Range Min-Max tree \cite{bpsoda10,bpalx10}, that we use
to support fast operations on balanced parentheses, improving
both in space and time on the existing
implementations \cite{bpalx10}. Our experimental results are discussed in
Section~\ref{sec:experimental-analys}, where our implementations compare very
favorably to some of the best implementations. We provide the source
code at \url{http://github.com/ot/path_decomposed_tries} for the
reader interested in further comparisons.

\subsection{Background and tools}
\label{sub:background-tools}
In the following we make extensive use of compacted tries and basic
succinct data structures.

\ttlpar{Compacted tries.} 
To fix the notation we recall quickly the definition of compacted
tries. We build recursively the trie in the following way. Basis: The
compacted trie of a single string is a node whose label is the
string. Inductive step: Given a nonempty string set $\mathcal{S}$, the
root of the tree is labeled with the longest common prefix $\alpha$
(possibly empty) of the strings in $\mathcal{S}$. For each character
$b$ such that the set $\mathcal{S}_b = \{\beta | \alpha b\beta \in
\mathcal{S}\}$ is nonempty, the compacted trie built on
$\mathcal{S}_b$ is attached to the root as a child. The edge is
labeled with the \emph{branching character} $b$.  The length of the
label $\alpha$ is also called the \emph{skip}, and denoted with
$\delta$. Unless otherwise specified, we will use \emph{trie} to
indicate a \emph{compacted trie} in the rest of the paper.

\ttlpar{Rank and Select operations.}
Given a bitvector $X$, we can define the following operations: 
$\Rank_b(i)$ returns the number of occurrences of bit $b
\in \{ \bitzero, \bitone\}$ in the first $i$ positions in $X$;
$\Select_b(i)$ returns the position of the $i$-th occurrence of bit $b$
in $X$. These operations can be supported in constant time by adding a 
negligible redundancy to the bitvector \cite{Clark98,jacobson89}.

\ttlpar{Elias-Fano encoding.} 
The \emph{Elias-Fano} representation \cite{elias74,fano71} is an
encoding scheme to represent a non-decreasing sequence of $m$ integers
in $[0, n)$ occupying $2m + m \left\lceil \log \frac n m \right\rceil
+ o(m)$ bits, while supporting constant-time access to the $i$-th
integer. 
The scheme is very simple and
elegant, and efficient implementations are described in
\cite{grossi05,sadaalx07,vigna08}.

\ttlpar{Balanced parentheses (BP).}  
In a sequence of balanced parentheses each open parenthesis $\paropen$
can be associated to its \emph{mate} $\parclose$. Operations
$\FindClose$ and $\FindOpen$ can be defined, which find the mate of
respectively an open and closed parenthesis. The sequences can be
represented as bitvectors, where $\bitone$ represents $\paropen$ and
$\bitzero$ represents $\parclose$, and by adding a negligible
redundancy it is possible to support the above defined operations in
constant or nearly-constant time \cite{jacobson89,munro97}.

\section{String Dictionaries}
\label{sec:stringdict}
In this section we describe an implementation of string dictionaries
using path-decomposed tries. A \emph{string dictionary} is a data
structure on a string set $\mathcal{S} \subset \Sigma^*$ that supports
the following operations:
\begin{itemize}
\shrinkitems
\item $\Lookup(s)$ returns $-1$ if $s \not\in \mathcal{S}$ or an
  unique identifier in $[0, |\mathcal{S}|)$ otherwise.
\item $\Access(i)$ retrieves the string with identifier $i$; note that
  $\Access(\Lookup(s)) = s$ if $s \in \mathcal{S}$.
\end{itemize}

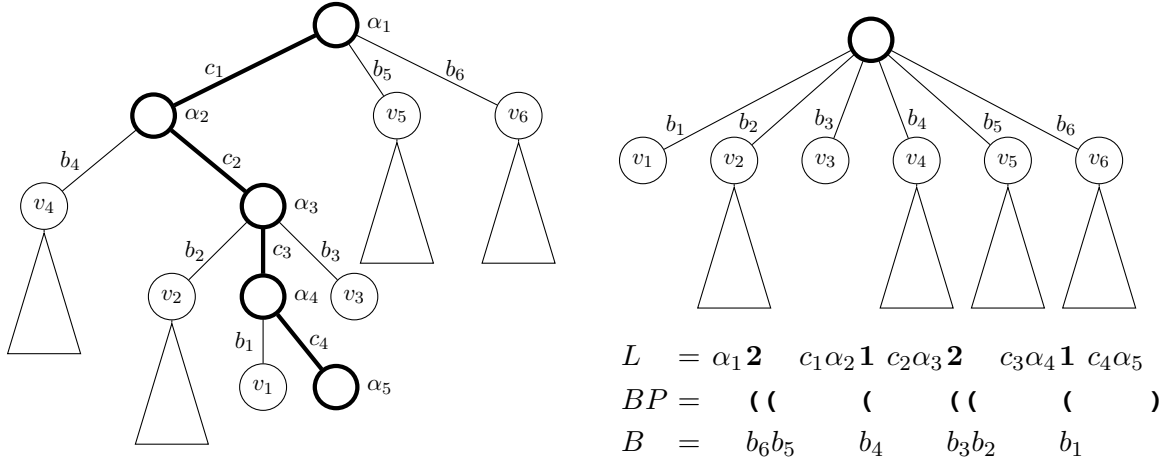
\begin{figure}[htbp]
  \begin{minipage}[b]{0.5\linewidth}
\begin{tikzpicture}[
  scale=0.8, transform shape,
  trienode/.style={circle, draw=black, fill=none, minimum width=0.7cm},
  heavy/.style={ultra thick},
  light/.style={thin},
  subtrie/.style={isosceles triangle, draw=black, shape border rotate=90,isosceles triangle stretches=true, minimum height=20mm,minimum width=12mm,inner sep=0,yshift={-20mm}},
  level distance=1.5cm,
  level/.style={sibling distance=2cm},
  level 2/.style={sibling distance=1.8cm},
  level 3/.style={sibling distance=1.5cm},
  level 4/.style={sibling distance=1.2cm}
  ]

  \node[heavy, trienode, label=right:$\alpha_1$] {}
  child {
    node[heavy, trienode, label=right:$\alpha_2$] {}
    child {
      node[trienode, light] {$v_4$}
      child [level distance=0]{ 
        node[subtrie, light] {}
      }
      edge from parent[left, light, inner sep=0.3cm] node {$b_4$}
    }
    child { node {} edge from parent[draw=none] } %
    child {
      node[trienode, heavy, label=right:$\alpha_3$] {}
      child {
        node[trienode, light] {$v_2$}
        child [level distance=0]{ 
          node[subtrie, light] {}
        }
        edge from parent[left, light, inner sep=0.2cm] node {$b_2$}
      }
      child {
        node[trienode, heavy, label=right:$\alpha_4$] {}
        child { node {} edge from parent[draw=none] } %
        child {
          node[trienode, light] {$v_1$}
          edge from parent[left, light] node {$b_1$}
        }
        child {
          node[trienode, heavy, label=right:$\alpha_5$] {}
          edge from parent[right, heavy] node {$c_4$}
        }
        edge from parent[right, heavy] node {$c_3$}
      }
      child {
        node[trienode, light] {$v_3$}
        edge from parent[right, light, inner sep=0.2cm] node {$b_3$}
      }
      edge from parent[right, heavy, inner sep=0.2cm] node {$c_2$}
    }
    edge from parent[left, heavy, inner sep=0.3cm] node {$c_1$}
  }
  child { node {} edge from parent[draw=none] } %
  child {
    node[trienode] {$v_5$}
    child [level distance=0]{ 
      node[subtrie] {}
    }
    edge from parent node[right, inner sep=0.1cm] {$b_5$}
  }
  child {
    node[trienode] {$v_6$}
    child [level distance=0]{ 
      node[subtrie] {}
    }
    edge from parent node[right, inner sep=0.3cm] {$b_6$}
  }
  ;

\end{tikzpicture}
    \vspace{0.4cm}
  \end{minipage}
  \begin{minipage}[b]{0.5\linewidth}
\begin{tikzpicture}[
  scale=0.8, transform shape,
  trienode/.style={circle, draw=black, fill=none, minimum width=0.7cm},
  heavy/.style={ultra thick},
  light/.style={thin},
  subtrie/.style={isosceles triangle, draw=black, shape border rotate=90,isosceles triangle stretches=true, minimum height=20mm,minimum width=12mm,inner sep=0,yshift={-20mm}},
  level distance=2cm,
  level/.style={sibling distance=1.5cm},
  ]

  \node[heavy, trienode] {}
  child {
    node[trienode, light] {$v_1$}
    edge from parent[near end, left, light, inner sep=0.4cm] node {$b_1$}
  }
  child {
    node[trienode, light] {$v_2$}
    child [level distance=0]{ 
      node[subtrie, light] {}
    }
    edge from parent[near end, left, light, inner sep=0.3cm] node {$b_2$}
  }
  child {
    node[trienode, light] {$v_3$}
    edge from parent[near end, left, light, inner sep=0.1cm] node {$b_3$}
  }
  child {
    node[trienode, light] {$v_4$}
    child [level distance=0]{ 
      node[subtrie, light] {}
    }
    edge from parent[near end, right, light, inner sep=0.1cm] node {$b_4$}
  }
  child {
    node[trienode, light] {$v_5$}
    child [level distance=0]{ 
      node[subtrie, light] {}
    }
    edge from parent[near end, right, light, inner sep=0.3cm] node {$b_5$}
  }
  child {
    node[trienode, light] {$v_6$}
    child [level distance=0]{ 
      node[subtrie, light] {}
    }
    edge from parent[near end, right, light, inner sep=0.4cm] node {$b_6$}
  }
  ;

\end{tikzpicture}
    \vspace{0.2cm}\\
\begin{tikzpicture}[
  scale=0.8, transform shape,
  every node/.style={anchor=base west}
  ]
  \matrix[
  matrix of math nodes, 
  inner xsep=0.1ex
  ] {
    L &[1mm] = & \alpha_1 & {\bf 2} & c_1 \alpha_2 & {\bf 1} & c_2 \alpha_3 & {\bf 2} & c_3 \alpha_4 & {\bf 1} & c_4 \alpha_5  \\
    BP &[1mm] = & &  \paropen\paropen & & \paropen & & \paropen\paropen & & \paropen & & \parclose \\
    B &[1mm] = & &  b_6 b_5 & & b_4 & & b_3 b_2 & & b_1\\
  };
\end{tikzpicture}
  \end{minipage}
  \caption{Path decomposition of a  trie. The $\alpha_i$ denote
    the labels of the trie nodes, $c_i$ and $b_i$ the branching
    characters (depending on whether they are on the path or not).}
  \label{fig:pathdec}
\end{figure}

\ttlpar{Path decomposition.}
Our string dictionaries, inspired by the approach described in
\cite{pods08}, are based on \emph{path decompositions} of the trie
built on $\mathcal{S}$ (recall that we use \emph{trie} to indicate 
\emph{compacted trie} in the rest of the paper).  A path decomposition
$\mathcal{T}^c$ of a trie $\mathcal{T}$ is a tree where each node in
$\mathcal{T}^c$ represents a path in $\mathcal{T}$. It is defined
recursively in the following way: a root-to-leaf path in $\mathcal{T}$
is chosen and represented by the root node in $\mathcal{T}^c$. The
same procedure is applied recursively to the sub-tries hanging off the
chosen path, and the obtained trees become the children of the root.
Note that in the above procedure the order of the decomposed sub-tries as
children of the root is arbitrary. Unlike \cite{pods08}, that arranges
the sub-tries in \emph{lexicographic order}, we arrange them in
\emph{bottom-to-top left-to-right} order since this simplifies the
traversal. Figure~\ref{fig:pathdec} shows a path in $\mathcal{T}$ and its resulting node in $\mathcal{T}^c$.

There is a one-to-one correspondence on the paths: root-to-\emph{node}
paths in $\mathcal{T}^c$ correspond to root-to-\emph{leaf} paths in
the trie $\mathcal{T}$, hence to strings in $\mathcal{S}$. This
implies also that $\mathcal{T}^c$ has exactly $|\mathcal{S}|$ nodes,
and the height of $\mathcal{T}^c$ can not be larger than that of
$\mathcal{T}$.
Different strategies in choosing the paths in the decomposition give
rise to different properties. We describe two such strategies.
\begin{itemize}
\shrinkitems
\item \emph{Leftmost path}: Always choose the leftmost child.
\item \emph{Heavy path}: Always choose the \emph{heavy} child,
  i.e. the one whose sub-trie has the most leaves (arbitrarily breaking
  ties). This is the strategy adopted in \cite{pods08} and borrowed
  from \cite{SleatorTarjanDynamicTrees}.
\end{itemize}

\begin{observation}
  If the leftmost path is used in the path decomposition, the
  depth-first order of the nodes in $\mathcal{T}^c$ is equal to the
  depth-first order of their corresponding leaves in
  $\mathcal{T}$. Hence if $\mathcal{T}$ is lexicographically ordered,
  so is $\mathcal{T}^c$. We call it a \emph{lexicographic path
    decomposition}.
\end{observation}

\begin{observation}
  \label{obs:centroid}
  If the heavy path is used in the path decomposition, the height of
  the resulting tree is bounded by $O(\log |\mathcal{S}|)$. We call
  such a decomposition a \emph{centroid path decomposition}.
\end{observation}

The two strategies enable a time/functionality trade-off: a
lexicographic path decomposition guarantees that the indices returned
by the $\Lookup$ are lexicographic, at cost of a potentially linear
height of the tree (but never higher than the  trie). On the
other hand, if the order of the indices is irrelevant, the centroid
path decomposition gives logarithmic guarantees.%
\footnote{In \cite{pods08} the authors show how to have
lexicographic indices in a centroid path-decomposed trie, using
secondary support structures and arranging the nodes in a different
order. The navigational operations are noticeably more
complex, and require more powerful primitives on the underlying
succinct tree, in particular for $\Access$.}

We exploit a crucial property of path decompositions: since each node
in $\mathcal{T}^c$ corresponds to a node-to-leaf path in
$\mathcal{T}$, the concatenation of the labels in the node-to-leaf
path corresponds to a suffix of a string in $\mathcal{S}$. To simulate
a traversal of $\mathcal{T}$ using $\mathcal{T}^c$ we only need to
scan sequentially character-by-character the label of each node until
we find the needed child node. Hence, any representation of the labels
that supports \emph{sequential access} (simpler than random access) is
sufficient. Besides being cache-friendly, as we will see in the next
section, this allows an efficient compression of the labels.

\ttlpar{Trie representation.} 
We represent the path-decomposed trie with three sequences (see
Figure~\ref{fig:pathdec}, containing an example for the root node):
\begin{itemize}\shrinkitems
\item The bitvector $BP$ encodes the trie topology using DFUDS
  \cite{dfuds}: each node is represented as a run of $\paropen$s of
  length the degree of the node, followed by a single $\parclose$; the
  node representations are then concatenated in depth-first order.
\item The array $B$ contains the branching characters of each node:
  they are written in reverse order per node, and then concatenated in
  depth-first order. Note that the branching characters are in
  one-to-one correspondence with the $\paropen$s of $BP$.
\item The sequence $L$ contains the \emph{labels} of each node. We
  recall that each label represents a \emph{path} in the trie. We
  encode the path augmenting the alphabet $\Sigma$ with $|\Sigma| - 1$
  special characters, $\Sigma' = \Sigma \cup \{{\bf 1}, {\bf 2},
  \dots, {\bf |\Sigma| - 1}\}$, alternating the label and the
  branching char of each node in the path with the number of sub-tries
  hanging off that node, encoded with the new special characters. We
  concatenate the representations of the labels in depth-first order
  in the sequence $L$, so that each label is in correspondence with a
  $\parclose$ in $BP$. Note that the labels are represented in a
  larger alphabet; we will show later how to encode them. Also, since the
  label representations are variable-sized, we encode their endpoints
  using an Elias-Fano sequence.
\end{itemize}

\ttlpar{Trie operations.} 
To implement $\Lookup$ we start from the root and begin scanning its
label. If the character is a special character, we add it to an
accumulator, otherwise we check for a mismatch with the string at the
current position. When there is a mismatch, the accumulator indicates
the range of children of the root (and thus of branching characters)
that branch from that point in the path in the original trie. Hence we
can find the right branching character (or conclude that that there is
none, i.e. the string was not in the set) and then the child where to
jump. We then proceed recursively until the string is fully traversed
or it cannot be extended further: the index returned is the value of
$\Rank_{\parclose}$ for the final node in the former case (i.e.\mbox{}
the depth-first index of that node), or $-1$ in the latter
case. Note that it is possible to avoid all the $\Rank$ calls needed
to access $L$ and $B$ by using the standard trick of double-counting,
i.e. exploiting the observation that between two mates there is an equal
number of $\paropen$s and $\parclose$s.

$\Access$ is performed similarly but in a bottom-up fashion. The node
position is obtained from the index through a $\Select_{\parclose}$,
then the path is reconstructed jumping to the parent until the node is
reached. Since we know for each node which child we came from, we can
scan its label until the sum of special characters encountered exceeds
the child index. The normal characters seen during the scan are
appended to the string to be returned.

\ttlpar{Time complexity.}
For the $\Lookup$, for each node in the traversal we perform a
sequential scan of the labels and a binary search on the branching
character. If the pattern has length $p$, we can never see more than
$p$ special characters during the scan. Hence if we assume
constant-time $\FindClose$ and Elias-Fano retrieval the total number
of operations is $O(p + h \log |\Sigma|)$, while the number of random
memory accesses is bounded by $O(h)$, where $h$ is the height of the
path decomposition tree. The $\Access$ is symmetric except that the
binary search is not needed and $p \geq h$, so the number of
operations is bounded by $O(p)$ where $p$ is the length of the
returned string. Again, the number of random memory accesses is
bounded by $O(h)$.

\ttlpar{Labels encoding and compression.} 
As previously mentioned, we need only to \emph{scan} sequentially the
label of each node, so we can use any encoding
that supports sequential scan with a constant amount of work per
character.
In the uncompressed trie, as a baseline, we simply use a vbyte encoding
\cite{vbyte}. Since most bytes in the datasets do not exceed $127$ as
a value, there is no noticeable space overhead. For a less sparse
alphabet, more sophisticated encodings can be used.

The freedom in choosing the encoding allows us to explore other
trade-offs. We take advantage of this to \emph{compress the labels},
with an almost negligible overhead in the operations.

We adopt a simple dictionary compression scheme for the labels: we
choose a static dictionary of variable-sized words (that can be drawn from any
alphabet) that will be stored along the tree explicitly, such
that the overall size of the dictionary is bounded by a given
parameter (constant) $D$. The node labels are then parsed into words
of the dictionary, and the words are sorted according to their
frequency in the parsing: a code is assigned to each
word in decreasing order of frequency, so that more frequent words
have smaller codes. The codes are then encoded using some
variable-length integer encoding; we use vbyte to favor
performance.
To decompress the label, we scan the codes and for each code we scan
the word in the dictionary, hence each character requires a constant
amount of work. 

We remark that the decompression algorithm is completely agnostic of
how the dictionary was chosen and how the strings are parsed. For example, 
domain knowledge about the data could be exploited; in texts, the most
frequent words would probably be a good choice. %

Since we are looking for a general-purpose scheme, we used a modified
version of the approximate Re-Pair \cite{repair} described in
\cite{navarrorepair}: we initialize the dictionary to the alphabet
$\Sigma$ and scan the string to find the $k$ most frequent pairs of
codes. Then we select all the pairs whose corresponding substrings fit
in the dictionary and substitute them in the sequence. We then iterate
until the dictionary is filled (or there are no more repeated pairs).
From this we obtain simultaneously the dictionary and the parsing.
To allow the labels to be accessed independently, we take care that no
pairs are formed on label boundaries, as done in \cite{csd11}.

Note that in principle our dictionary representation is less
space-efficient than plain Re-Pair, where the words are represented
recursively as pairing rules. However accessing a single character
from a recursive rule has a cost dependent on the rule tree height, so
it would fail our requirement of constant amount of work per decoded
character.

\ttlpar{Implementation notes.} 
For the $BP$ vector we use the Range Min tree described in
Section~\ref{sec:rmtree}. $\Rank$ is supported using the
\texttt{rank9} structure described in \cite{vigna08}, while $\Select$
is implemented through a one-level hinted binary search. The search for
the branching character is replaced by a linear search, which for the
cardinalities considered is actually \emph{faster} in practice. 
The dictionary is represented as the concatenation of the
words encoded in $16$-bit characters to fit the larger alphabet
$\Sigma' = [0, 511)$. The dictionary size bound $D$ is chosen to be
$2^{16}$, so that the word endpoints can be encoded in $16$-bit
pointers. The small size of the dictionary makes also more likely that
(at least the most frequently accessed part of) it is kept in cache.

\section{Monotone Minimal Perfect Hash for Strings}
\label{sec:hollow}

Minimal perfect hash functions map a set of strings $\mathcal{S}$
bijectively into $[0, |\mathcal{S}|)$. \emph{Monotone} minimal perfect
hash functions \cite{monotonehashsoda} (or \emph{monotone hashes})
also require that the mapping preserves the lexicographic order of the
strings (not to be confused with generic order-preserving hashing).
We remark that, as for standard minimal hash functions, the $\Lookup$
can return any number on strings outside of $\mathcal{S}$, hence the
data structure does not have to \emph{store} the string set.

The hollow trie \cite{monotonehash09} is a particular instance of
monotone hash. It consists in a binary trie on $\mathcal{S}$, of which
only the trie topology and the skips of the internal nodes are stored,
in succinct form. To compute the hash value of a string~$x$, a \emph{blind
  search} is performed: the trie is traversed matching only the
branching characters (bits, in this case) of~$x$. If $x \in
\mathcal{S}$, the leaf reached is the correct one, and its unique
identifier in $[0, |\mathcal{S}|)$ is returned; otherwise, it has the
longest prefix match with~$x$, useful in some applications.

The cost of unbalancedness for hollow tries is even larger than that
for normal tries: since the strings over $\Sigma$ have to be converted
to a binary alphabet, the height is potentially multiplied by $O(\log
|\Sigma|)$ with respect to that of a trie on $\mathcal{S}$. The
experiments in \cite{monotonehash09} show indeed that the data
structure is not practical compared to other monotone hashes
analyzed in that paper.

\begin{figure}[htbp]
 \hspace{1cm} %
  \begin{minipage}[b]{0.5\linewidth}
\begin{tikzpicture}[
  scale=0.8, transform shape,
  trienode/.style={circle, draw=black, fill=none, minimum width=0.7cm},
  heavy/.style={ultra thick},
  light/.style={thin},
  subtrie/.style={isosceles triangle, draw=black, shape border rotate=90,isosceles triangle stretches=true, minimum height=15mm,minimum width=12mm,inner sep=0,yshift={-15mm}},
  level distance=1.5cm,
  level/.style={sibling distance=2cm},
  level 2/.style={sibling distance=1.5cm},
  level 3/.style={sibling distance=1.5cm},
  level 4/.style={sibling distance=1.2cm}
  ]

  \node[heavy, trienode, label=right:$\delta_1$] {}
  child {
    node[heavy, trienode, label=right:$\delta_2$] {}
    child {
      node[trienode, light] {$v_1$}
      child [level distance=0]{ 
        node[subtrie, light] {}
      }
      edge from parent[left, light, inner sep=0.3cm] node {\bitzero}
    }
    child { node {} edge from parent[draw=none] } %
    child {
      node[trienode, heavy, label=right:$\delta_3$] {}
      child {
        node[trienode, light] {$v_2$}
        edge from parent[left, light, inner sep=0.2cm] node {\bitzero}
      }
      child {
        node[trienode, heavy, label=right:$\delta_4$] {}
        child { node {} edge from parent[draw=none] } %
        child {
          node[trienode, heavy] {}
          edge from parent[left, heavy] node {\bitzero}
        }
        child {
          node[trienode, light] {$v_3$}
          edge from parent[right, light, inner sep=0.3cm] node {\bitone}
        }
        edge from parent[right, heavy] node {\bitone}
      }
      edge from parent[right, heavy, inner sep=0.2cm] node {\bitone}
    }
    edge from parent[left, heavy, inner sep=0.3cm] node {\bitzero}
  }
  child { node {} edge from parent[draw=none] } %
  child {
    node[trienode] {$v_4$}
    child [level distance=0]{ 
      node[subtrie] {}
    }
    edge from parent node[right, inner sep=0.3cm] {\bitone}
  }
  ;

\end{tikzpicture}
    \vspace{0.5cm}
  \end{minipage}
  \begin{minipage}[b]{0.5\linewidth}
\begin{tikzpicture}[
  scale=0.8, transform shape,
  trienode/.style={circle, draw=black, fill=none, minimum width=0.7cm},
  heavy/.style={ultra thick},
  light/.style={thin},
  subtrie/.style={isosceles triangle, draw=black, shape border rotate=90,isosceles triangle stretches=true, minimum height=15mm,minimum width=12mm,inner sep=0,yshift={-15mm}},
  level distance=2cm,
  level/.style={sibling distance=1.5cm},
  ]

  \node[heavy, trienode] {}
  child {
    node[trienode, light] {$v_1$}
    child [level distance=0]{ 
      node[subtrie, light] {}
    }
    edge from parent[near end, left, light, inner sep=0.3cm] node {\bitzero}
  }
  child {
    node[trienode, light] {$v_2$}
    edge from parent[near end, left, light, inner sep=0.1cm] node {\bitzero}
  }
  child {
    node[trienode, light] {$v_3$}
    edge from parent[near end, right, light] node {\bitone}
  }
  child {
    node[trienode, light] {$v_4$}
    child [level distance=0]{ 
      node[subtrie, light] {}
    }
    edge from parent[near end, right, light, inner sep=0.3cm] node {\bitone}
  }
  ;

\end{tikzpicture}
\begin{tikzpicture}[
  scale=0.8, transform shape,
  every node/.style={anchor=base west},
  arr/.style={->,dashed,in=90,out=270,looseness=.3},
  ]
  \matrix (A) [
  matrix of math nodes, 
  inner xsep=0.1ex
  ] {
    \Lhigh &[1mm] = & \bitzero^{|\delta_1|} & \bitone & \bitzero^{|\delta_2|} & \bitone & \bitzero^{|\delta_3|} & \bitone & \bitzero^{|\delta_4|} & \bitone \\
    \Llow &[1mm] = & \delta_1 & \bitzero & \delta_2 & \bitone & \delta_3 & \bitone & \delta_4 & \bitzero \\[0.5cm]
    BP &[1mm] = & & \paropen & & \paropen & & \paropen & & \paropen & \parclose \\
  };
  
  \draw [arr] (A-2-4.south) to (A-3-4.north);
  \draw [arr] (A-2-6.south) to (A-3-10.north);
  \draw [arr] (A-2-8.south) to (A-3-8.north);
  \draw [arr] (A-2-10.south) to (A-3-6.north);
\end{tikzpicture}
  \end{minipage}
  \caption{Path decomposition of a hollow trie. The $\delta_i$ denote the skips.}
  \label{fig:binpathdec}
\end{figure}
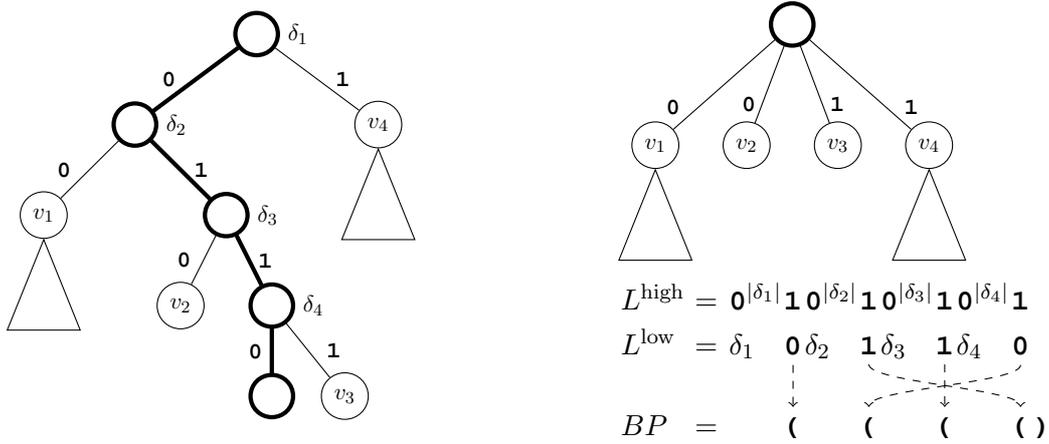

\ttlpar{Path decomposition with lexicographic order.}
To tackle their unbalancedness, we apply the centroid path
decomposition idea to hollow tries. The construction presented in
Section~\ref{sec:stringdict} cannot be used directly, because we want 
to both preserve the lexicographic ordering of the strings \emph{and} 
guarantee the logarithmic height. However
both the binary alphabet and the fact that we do not need the
$\Access$ operation come to the aid.
First, inspired again by \cite{pods08}, we arrange the sub-tries in
\emph{lexicographic} order. This means that the sub-tries on the
\emph{left} of the path are arranged top-to-bottom, and precede all
those on the \emph{right} which are arranged bottom-to-top. In the
path decomposition tree we call \emph{left} children the ones
corresponding to sub-tries hanging off the left side of the path and
\emph{right} children the ones corresponding to those hanging on the
right side. Figure~\ref{fig:binpathdec} shows the new ordering.

We now need a small change in the heavy path strategy: instead of
breaking ties arbitrarily, we choose the \emph{left} child. We call
this strategy \emph{left-biased heavy path}, which gives the
following.
\begin{observation}
  \label{obs:rightchild}
  Every node-to-leaf left-biased heavy path in a binary trie ends with
  a left turn. Hence, every internal node of the resulting path
  decomposition has at least one right child.
\end{observation}

\ttlpar{Trie representation.} 
The bitvector $BP$ is defined as in Section~\ref{sec:stringdict}. The
label associated with each node is the sequence of skips interleaved
with directions taken in the centroid path, excluding the leaf skip,
as in Figure~\ref{fig:binpathdec}. Two aligned bitvectors $\Lhigh$ and $\Llow$
are used to represent the labels using an encoding inspired by
$\gamma$ codes: the skips are incremented by one (to exclude $0$ from
the domain) and their binary representations (without the leading $\bitone$) are interleaved with the
path directions and concatenated in $\Llow$. $\Lhigh$ consists 
$\bitzero$ runs of length corresponding to the 
lengths of the binary representations of the skips, followed by
$\bitone$s, so that the endpoints of (skip, direction) pair encodings in $\Llow$
correspond to the $\bitone$s in $\Lhigh$. Thus a $\Select$ directory on
$\Lhigh$ enables random access to the (skip, direction) pairs
sequence. The labels of the node are concatenated in depth-first
order: the $\paropen$s in $BP$ are in one-to-one correspondence
with the (skip, direction) pairs.

\ttlpar{Trie operations.} 
As in Section~\ref{sec:stringdict}, a trie traversal is simulated on
the path decomposition tree. In the root node, the (skip, direction)
pairs sequence is scanned (through $\Lhigh$ and $\Llow$): during the
scan the number of left and right children passed by is kept; when a
mismatch in the string is found, the search proceeds in the
corresponding child. Because of the ordering of the children, if the
mismatch leads to a left child the child index is the number of left
children seen in the scan, while if it leads to a right child it is
the node degrees minus the number of right children seen (because the
latter are stored from right to left). The search proceeds recursively
until the string's characters are consumed.

When the search ends, the depth-first order of the node found is not yet the
number we are looking for: all the ancestors where we turned
\emph{left} come before the found node in depth-first but \emph{after} in the
lexicographic order. Besides, if the found node is not a leaf, all the
strings in the left sub-tries of the corresponding path are
lexicographically smaller than the current string.
It is easy to fix these issues: during the traversal we can count the
number of left turns and subtract that from the final index. To
account for the left sub-tries, using Observation~\ref{obs:rightchild}
we can count the number of their leaves by jumping to the first right
child with a $\FindClose$: the number of nodes skipped in the jump is
equal to the number of leaves in the left sub-tries of the node.

\ttlpar{Time complexity.}
The running time of the $\Lookup$ can be analyzed with a similar
argument to that of the $\Lookup$ of
Section~\ref{sec:stringdict}: during the scan there cannot be more
skips than the pattern length; besides there is no binary
search. Hence the number of operations is $O(\min(p, h))$, 
while the number of random
memory accesses is bounded by $O(h)$.

\ttlpar{Implementation notes.} 
To support the $\Select$ on $\Lhigh$ we
use a variant of the \emph{darray} \cite{sadaalx07}: since the $\bitone$s in
the sequence are at most $64$ bits apart, we can bound the size of the
blocks so that we do not need the overflow vector (called $S_l$ in
\cite{sadaalx07}).

\section{Balanced Parentheses: The Range Min Tree}
\label{sec:rmtree}

In this section we describe the data structure supporting $\FindClose$
and $\FindOpen$. As it is not restricted to tries, we believe it is of
independent interest.

We begin by discussing the Range Min-Max tree \cite{bpsoda10}, which
is a succinct data structure to support operations on balanced
parentheses in $O(\log n)$ time. It was shown in \cite{bpalx10} that
it is very efficient in practice. Specifically, it is a data structure
on $\{-1, 0, +1\}$ sequences that supports the forward search
$\fwdsearch(i, x)$: given a position $i$ and a target value $x$,
return the first position $j > i$ such that the sum of the values in
the sequence between $i$ and $j$ is equal to $x$. The application to
balanced parentheses is straightforward: if the sequence takes value
$+1$ on open parentheses on $-1$ on closed parentheses, $\FindClose(i)
= \fwdsearch(i, 0)$. In other words, it is the first position with
zero \emph{excess}, defined as the difference between the number of
open and close parentheses up to the given position.  Backwards search
is defined similarly for $\FindOpen$.

The data structure is defined as follows: the sequence is divided in
\emph{blocks} of the same size and a \emph{tree} is formed over the
blocks, storing the minimum $m_i$ and the maximum $M_i$ of the
sequence cumulative sum (the excess, for balanced parentheses) for
each block $i$ in the leaves, and for the sub-trees in the nodes. The
forward search traverses the tree to find the first block $j$ after
$i$ where the target value $x$ is between $m_j$ and $M_j$. Since the
sequence is $\{-1, 0, +1\}$, block $j$ contains all the intermediate
values between $m_j$ and $M_j$, and so it must contain $x$. A linear
search is then performed within the block (usually through lookup
tables).

We fit the above data structure to support only $\FindOpen$ and
$\FindClose$, thus reducing both the space requirement and the time
performance. We list our two modifications.

\ttlpar{Halving the tree space.}  
We discard the maxima and store only the minima, and call the
resulting tree \emph{Range Min tree}. During the block search, we only
check that target value is greater than the block minimum. The
following lemma guarantees that the forward search is correct. A
symmetric argument holds for the backwards search.
\begin{lemma}
  In a balanced parentheses sequence, the Range Min tree forward
  search for $x=0$ finds the same block as the Range Min-Max tree.
\end{lemma}
\begin{Proof}
  Since the Min search is a relaxation of the Min-Max search, the
  block $j'$ returned by the search in the Min tree must precede the
  block $j$ found by Min-Max search, i.e. $j' \leq j$.
  Suppose by contradiction that $j' < j$. Since the sequence of
  parentheses is balanced, all the positions between two mates have
  excess greater than the excess of the opening parenthesis. Then
  $M_{j'}$ is greater than the excess of the opening parenthesis,
  which is the target value. Hence $j'$ is a valid block for the
  forward search in the Min-Max tree, but since $j' < j$ we have a contradiction. 
\end{Proof}

\ttlpar{Broadword in-block search.} 
The in-block search performance is crucial as it is the inner loop of
the search. In practical implementations it is usually performed
byte-by-byte with a lookup table that contains the solution for each
possible byte and excess.  This involves many branches and accesses to
a fairly big lookup tables for each byte. Supposing instead that we
know which byte contains the closing parenthesis, we can then use the
lookup table only on that byte. %

To find that byte we can use the same trick as the
Range Min: the first byte with min-excess smaller than the target
excess contains the closing parenthesis. We find it with an hybrid
lookup table/broadword approach.

We divide the block into machine words. For each word $w$ we compute the
word $m_8$ where the $i$-th byte contains the min-excess of
the $i$-th byte in $w$ with inverted sign, so that it is non-negative. This is achieved through a pre-computed lookup
table which contains the min-excess for each possible byte. At the
same time we compute the byte counts $c_8$ of $w$, where
the $i$-th byte contains the number of $\bitone$s of the $i$-th byte of $w$, 
using the algorithm described in \cite{knuthtaocp}. 

Using the equality $\Excess(i) = 2\cdot
\Rank_\paropen(i) - i$ we can easily compute the excess for each byte
of $w$: if $e_w$ is the excess at the starting position of $w$, the
word $e_8$ whose $i$-th byte contains the excess of the $i$-th byte of
$w$ can be obtained through the following formula:\footnote{%
A subtlety is needed here for a correct search: the excess can be negative, hence the carry
in the subtraction corrupts the bytes after the first byte that
contains the zero. However, this means that the word \emph{contains}
the solution, and the closing parenthesis is in the byte that precedes
the one where the sampled excess goes negative.
}
\begin{equation*}
  e_8 = (e_w + ((2 * c_8 - \texttt{0x...08080808}) \texttt{ <{}< } 8)) * \texttt{0x...01010101} \text{.}
\end{equation*}

Now we have all we need: the closing parenthesis is in the byte where
the excess function crosses the zero, in other words in the byte whose
excess added to the min-excess is smaller than zero. Hence we are
looking for the first byte position in which $e_8$ is smaller than
$m_8$ (recall that the bytes in $m_8$ are negated). This can be done using the $\leq_8$ operation described in
\cite{knuthtaocp} to compute a mask $l_8 = e_8 \leq_8 m_8$, where the
$i$-th byte is $1$ if and only if the $i$-th byte of $e_8$ is smaller
than the $i$-th byte of $m_8$. If the $l_8$ is zero, the word does not
contain the closing parenthesis; otherwise, an LSB operation quickly
returns the index of the byte containing the solution. The same
algorithm can be applied symmetrically for the $\FindOpen$.

All in all, we performed $8$ lookups from a very small table, a few
arithmetic operations and one single branch (to check whether the word
contains the solution or not). In our experiments, the approach
described above results in $\approx 30\%$ faster operations in
tree-traversal benchmarks, with respect to byte-by-byte search. 
A similar improvement occurs in our trie
implementations.

\section{Experimental Analysis}
\label{sec:experimental-analys}
In this section we discuss a series of experiments we performed on
both real-world and synthetic data. We performed several tests both to
collect statistics that show how our path decompositions give an
algorithmic advantage over standard tries, and to benchmark the
implementations comparing them with other practical
data structures.

\ttlpar{Setting.} 
The experiments were run on a 64-bit 2.53GHz Core i7 processor with 8MB
L3 cache and 24GB RAM, running Windows Server 2008 R2. All the C++
code was compiled with MSVC 10, while for Java we used the Sun JVM 6.

\ttlpar{Datasets.} 
The tests were run on the following datasets.
\begin{itemize}\shrinkitems
\item \texttt{enwiki-titles} (163MiB, 8.5M strings): All the page
  titles from English Wikipedia.
\item \texttt{aol-queries} (224MiB, 10.2M strings): The queries in the
  AOL 2006 query log \cite{aolqueries}.
\item \texttt{uk-2002} (1.3GiB, 18.5M strings): The URLs of a 2002
  crawl of the \texttt{.uk} domain \cite{BCSU3}.
\item \texttt{webbase-2001} (6.6GiB, 114.3M strings): The URLs in
  the Stanford WebBase from \cite{lawdatasets}.
\item \texttt{synthetic} (1.4GiB, 2.5M strings): The set of strings
  $\bitnobf{d}^i \bitnobf{c}^j \bitnobf{b}^t \sigma_1\dots\sigma_k$ where $i$ and $j$ range in $[0,
  500)$, $t$ ranges in $[0, 10)$, $\sigma_i$ are all distinct (but
  equal for each string) and $k=100$. The resulting tries are very
  unbalanced, while the constant suffix $\sigma_1\dots\sigma_k$
  stresses the in-bucket search in bucketed front coding and the
  redundancy is not exploited by front coding and non-compressed
  tries. At the same time, the strings are extremely compressible.
\end{itemize}

\ttlpar{Average height.} 
Table~\ref{tab:stats} compares the average
height of plain tries with their path decomposition trees. In all the
real-world datasets the centroid path decomposition cause a $\approx 2$-$3$
times reduction in height compared to the standard compacted trie. The
gap is even more dramatic in hollow tries, where the binarization of
the strings causes a blow-up in height close to $\log |\Sigma|$, while
the centroid path-decomposed tree height is very small, actually much
smaller than $\log |\mathcal{S}|$. It is interesting to note that even
if the lexicographic path decomposition is unbalanced, it still
improves on the  trie, due to the higher fan-out of the
internal nodes. 

The synthetic dataset is a pathological case for tries, but the
centroid path-decomposition still maintains an extremely low average
height.

\begin{table}[tbp]
  \shrinktable
  \begin{center}
    { 
      \scriptsize
      \setlength{\tabcolsep}{1ex}
      \renewcommand\arraystretch{1} 
\begin{tabular}{lrrrrr}\toprule
&  \multicolumn{1}{c}{\scriptsize\texttt{enwiki-titles}}&\multicolumn{1}{c}{\scriptsize\texttt{aol-queries}}&\multicolumn{1}{c}{\scriptsize\texttt{uk-2002}}&\multicolumn{1}{c}{\scriptsize\texttt{webbase-2001}}&\multicolumn{1}{c}{\scriptsize\texttt{synthetic}} \\
\midrule
Compacted trie avg. height & 9.8 & 11.0 & 16.5 & 18.1 & 504.4 \\
Lex. avg. height & 8.7 & 9.9 & 14.0 & 15.2 & 503.5 \\
Centroid avg. height & 5.2 & 5.2 & 5.9 & 6.2 & 2.8 \\
\midrule
Hollow avg. height & 49.7 & 50.8 & 55.3 & 67.3 & 1005.3 \\
Centroid hollow avg. height & 7.9 & 8.0 & 8.4 & 9.2 & 2.8 \\
\bottomrule
\end{tabular}
    }
  \end{center}
  \caption{Average height: for tries the average height of
    the \emph{leaves} is considered, while for path-decomposed tries
    all the nodes are considered (see the comments after
    Observation~\ref{obs:centroid}).}
  \label{tab:stats}
\end{table}

\ttlpar{String dictionary data structures.} 
We compared the
performance of our implementations of path-decomposed tries to other
data structures. \emph{Centroid} and \emph{Centroid
  compr.} implement the centroid path-decomposed trie described in
Section~\ref{sec:stringdict}, in the versions without and with labels
compression. Likewise, \emph{Lex.} and \emph{Lex. compr.}  implement
the lexicographic version.

\emph{Re-Pair} and \emph{HTFC} are respectively the Re-Pair and
Hu-Tucker compressed Front Coding from
\cite{csd11}. %
For HTFC we chose bucket size $8$ as the best space/time
trade-off. Comparison with Front Coding is of particular interest as it is
one of the data structures generally preferred by the practitioners.

\emph{TX} is a popular open-source straightforward implementation of a
(non-compacted) trie that uses LOUDS \cite{jacobson89} to represent
the tree. The code can be downloaded
from \cite{tx}. We made some slight changes to avoid keeping all the
string set in memory during the construction

To measure the running times, we chose $1$ million random (and
randomly shuffled) strings from each dataset for the $\emph{Lookup}$
and $1$ million random indices for the $\Access$. Each test was
averaged on $10$ runs. The construction time was averaged on $3$ runs.

Re-Pair, HTFC and TX do not support files bigger than $2$GiB, so we
could not run the tests on \texttt{webbase-2001}. Also Re-Pair did not
complete the construction on \texttt{synthetic} in 6 hours, so we had
to kill the process.

\ttlpar{String dictionaries results.} 
The results of the tests can be
seen in Table~\ref{tab:results}. On all datasets our compressed tries
obtain the smaller space, except on \texttt{uk-2002} where they come a
close second. The centroid versions have also the fastest $\Lookup$
times, while the $\Access$ time is better for Re-Pair and occasionally
HTFC, whose time is although within $20$\% of that of the centroid
trie. TX is consistently the largest and slowest on all the datasets.

Maybe surprisingly, the lexicographic trie is not much slower than the
centroid trie for both $\Lookup$ and $\Access$. However in the
synthetic dataset the unbalanced tries are more than $20$ times slower
than the balanced ones. HTFC exhibits a less dramatic slowdown but still in
the order of $5$x on lookup compared to the centroid trie. Although
this behavior does not occur on our real-world datasets, it shows that
no assumptions can be made for unbalanced tries. For example in an
adversarial environment an attacker could exploit this weakness to
perform a denial of service attack.

We remark that the labels compression adds an almost negligible
overhead in both $\Lookup$ and $\Access$, due to the extremely simple
dictionary scheme, while obtaining a very good compression. 
Hence unless the construction time is a concern (in which case
other dictionary selection strategies can also be explored) 
it is always convenient to compress the labels.

\ttlpar{Monotone hash data structures.} 
For monotone hashes, we
compared our data structures with the implementations in
\cite{monotonehash09}. \emph{Centroid hollow} implements the centroid
path-decomposed hollow trie described in Section~\ref{sec:hollow}.
\emph{Hollow} is a reimplementation of the hollow trie of
\cite{monotonehash09} but using a Range Min tree in place of a
pioneer-based representation. \emph{Hollow (Sux)} and \emph{PaCo
  (Sux)} are two implementations from \cite{monotonehash09}; the first
is the hollow trie, the second an hybrid scheme: a Partially Compacted
trie is used to partition the keys into buckets, then each bucket is
hashed with an MWHC function. Among the structures in
\cite{monotonehash09}, PaCo gives the best trade-off between space and
lookup time. The implementations are freely available as part of the Sux project \cite{sux}.\footnote{To be fair we need to
say that Sux is implemented in Java while our structures are
implemented in C++. However, the recent developments of the Java
Virtual Machine have made the abstraction penalty gap smaller and
smaller. Low-level optimized Java (as the one in Sux) can be on par of
C++ for some tasks, and no slower than 50\% with respect to C++ for
most other tasks \cite{shootout}. We remark that the hollow trie
construction is actually \emph{faster} in the Sux version than in
ours, although the algorithm is very similar.}

To measure construction time and lookup time we adopted the same
strategy as for string dictionaries. For Sux, as suggested in
\cite{monotonehash09}, we performed $3$ runs of lookups before
measuring the lookup time, to let the JIT warm up and optimize the
generated code. 

\ttlpar{Monotone hash results.} 
Table~\ref{tab:results} shows the
results for monotone hashes. On all real-world datasets the centroid
hollow trie is $\approx 2$-$3$ times faster than our implementation of
the hollow trie and $\approx 5$ times faster than the Sux
implementation. The centroid hollow trie is competitive with PaCo on
all datasets, while taking less space and with a substantially simpler
construction. The synthetic dataset in particular triggers the
pathological behavior on all the unbalanced structures, with
\emph{Hollow}, \emph{Hollow (Sux)} and \emph{PaCo} being respectively
$13$, $41$ and $5$ times slower than \emph{Centroid hollow}. 
Such a large performance gap suggests the same conclusion reached for string dictionaries:
if \emph{predictable} performance is needed, unbalanced structures should be avoided.

\begin{table}[tbp]\shrinktable
  \begin{center}
    { 
      \tiny \setlength{\tabcolsep}{0.7ex} %
      \renewcommand\arraystretch{1.2} %
\begin{tabular}{lrrrr|rrrr|rrrr|rrrr|rrrr}\toprule
&  \multicolumn{4}{c}{\scriptsize\texttt{enwiki-titles}}&\multicolumn{4}{c}{\scriptsize\texttt{aol-queries}}&\multicolumn{4}{c}{\scriptsize\texttt{uk-2002}}&\multicolumn{4}{c}{\scriptsize\texttt{webbase-2001}}&\multicolumn{4}{c}{\scriptsize\texttt{synthetic}} \\
&  \multicolumn{4}{c}{161 bps}&\multicolumn{4}{c}{185 bps}&\multicolumn{4}{c}{621 bps}&\multicolumn{4}{c}{497 bps}&\multicolumn{4}{c}{4836 bps} \\
\midrule
& \multicolumn{20}{c}{String dictionaries}\\
\cmidrule(r){2-21}
&  ctps & c.ratio & lkp & acs & ctps & c.ratio & lkp & acs & ctps & c.ratio & lkp & acs & ctps & c.ratio & lkp & acs & ctps & c.ratio & lkp & acs \\
\midrule
Centroid compr. &  6.1 & 32.1\% & 2.5 & 2.6 & 7.9 & 31.5\% & 2.7 & 2.7 & 8.5 & 13.6\% & 3.8 & 4.9 & 7.8 & 13.5\% & 4.8 & 5.4 & 13.7 & 0.4\% & 4.2 & 13.5 \\
Lex. compr. &  6.4 & 31.9\% & 3.2 & 3.1 & 8.0 & 31.2\% & 3.8 & 3.6 & 8.4 & 13.5\% & 5.9 & 6.6 & 8.5 & 13.3\% & 7.3 & 7.7 & 109.2 & 0.4\% & 90.9 & 96.3 \\
Centroid &  1.8 & 53.6\% & 2.4 & 2.4 & 2.0 & 55.6\% & 2.4 & 2.6 & 2.3 & 22.4\% & 3.4 & 4.2 & 2.2 & 24.3\% & 4.3 & 5.0 & 8.4 & 17.9\% & 5.1 & 13.4 \\
Lex. &  2.0 & 52.8\% & 3.1 & 3.2 & 2.2 & 55.0\% & 3.5 & 3.5 & 2.7 & 22.3\% & 5.5 & 6.2 & 2.6 & 24.3\% & 7.0 & 7.4 & 102.8 & 17.9\% & 119.8 & 114.6 \\
Re-Pair \cite{csd11} &  60.0 & 41.5\% & 6.6 & 1.2 & 115.4 & 38.8\% & 7.3 & 1.3 & 326.4 & 12.4\% & 25.7 & 3.1 & - & - & - & - & - & - & - & - \\
HTFC \cite{csd11} &  0.4 & 43.2\% & 3.7 & 2.2 & 0.4 & 40.9\% & 3.8 & 2.2 & 0.9 & 24.4\% & 7.0 & 4.7 & - & - & - & - & 5.0 & 19.1\% & 22.0 & 18.0 \\
TX \cite{tx} &  2.7 & 64.0\% & 9.7 & 9.1 & 3.3 & 69.4\% & 11.9 & 11.3 & 5.7 & 30.0\% & 42.1 & 42.0 & - & - & - & - & 44.6 & 25.3\% & 284.3 & 275.9 \\
\midrule
& \multicolumn{20}{c}{Monotone hashes}\\
\cmidrule(r){2-21}
&  ctps & bps & lkp &  & ctps & bps & lkp &  & ctps & bps & lkp &  & ctps & bps & lkp &  & ctps & bps & lkp &  \\
\midrule
Centroid hollow &  1.1 & 8.40 & 2.7 &  & 1.2 & 8.73 & 2.8 &  & 1.5 & 8.17 & 3.3 &  & 1.5 & 8.02 & 4.4 &  & 8.6 & 9.96 & 11.1 &  \\
Hollow &  1.3 & 7.72 & 6.8 &  & 1.3 & 8.05 & 7.2 &  & 1.7 & 7.48 & 9.3 &  & 1.7 & 7.33 & 13.9 &  & 9.5 & 9.02 & 137.1 &  \\
Hollow (Sux \cite{sux}) &  0.9 & 7.66 & 14.6 &  & 1.0 & 7.99 & 16.6 &  & 1.1 & 7.42 & 18.5 &  & 0.9 & 7.27 & 22.4 &  & 4.3 & 6.77 & 462.7 &  \\
PaCo (Sux \cite{sux}) &  2.6 & 8.85 & 2.4 &  & 2.9 & 8.91 & 3.1 &  & 4.7 & 10.65 & 4.3 &  & 18.4 & 9.99 & 4.9 &  & 21.3 & 13.37 & 51.1 &  \\
\bottomrule
\end{tabular}
    }
  \end{center}
  \caption{Experimental results. \textbf{bps} is \emph{bits per string}, \textbf{ctps} 
    is the average \emph{construction time} per string, \textbf{c.ratio} is the \emph{compression ratio} 
    between the data structure and the original file sizes, 
    \textbf{lkp} is the average \emph{$\Lookup$} time and \textbf{acs} the average \emph{$\Access$} time. 
    All times are expressed in microseconds.
  }
  \label{tab:results}
\end{table}

\section{Conclusion and Future Work}

We have presented new succinct representation for tries that guarantee
low average height and enables the compression of the labels. Our
experimental analysis has shown that they obtain the best space
occupancy when compared to the state of the art, while maintaining
competitive access times. Moreover, they give the most
\emph{consistent} behavior among different (possibly synthetic)
datasets.
We have not considered alternatives for the dictionary selection
algorithm in labels compression; any improvement in that direction
would be beneficial to the space occupancy of our tries. We also plan
to consider other kinds of path decompositions (longest path, ladder,
etc.), which could enable other time/space/functionality trade-offs.

\section*{Acknowledgments}
We would like to thank the authors of \cite{csd11} for kindly providing the
source code for their algorithms.

\bibliographystyle{abbrv}
\bibliography{compressed_trie_paper}

\end{document}